1Alexander Greysukh# Internet-Connected Residential Water Leak Monitor

*Abstract*—A device for passive monitoring of slow water leaks, especially periodic leaks caused by faltering gaskets. The detection and notification solution is composed of an algorithm tuned specifically for bathroom leaks, motivated by a surprise water bill, and a hardware prototype connected to a serverless cloud component. The package is essentially a microservice for slow-leak detection and notification. Inexpensive commodity hardware is utilized to send alerts to users over Amazon Simple Notification Service (SNS).

*Index Terms*—home automation, water conservation, signal processing, IoT.## I. INTRODUCTION

Slow leaks left unattended cause significant water waste over long periods. Periodic leaks, e.g. caused by failing toilet tank flappers, are especially difficult to notice and typically precede a surprise water bill. Several projects and products that use magnetic, optical, and ultrasound sensors [1-3] cover the baseline reference material. High-end solutions are complex, usually integrated with a shut-off valve, require professional installation, and are orders of magnitude more expensive than this approach. Popular open-source-hardware microcontrollers and sensors are powerful enough for designing simple, inexpensive, and non-intrusive solutions. In a sense, this solution is a microservice - focusing on one problem and solving it efficiently. It can be a module in a smart home ecosystem or used independently.

Common residential water meters are equipped with a revolving piston, rotated by flowing water. Rotation frequency depends on the water flow and can be measured by sensing magnetic field fluctuations emitted by the meter's internal magnetic drive. The sensitivity of a magnetometer placed near a water meter is sufficient for receiving a reliable signal. Real-time processing and statistical analysis allow for detecting leaks in a noisy signal. All the computations are conducted on-board, making the device operationally independent. Alert delivery is delegated to the Amazon cloud.

Piston frequency in the 0.01-0.4 Hz range corresponds to approximately a 1-20 Gal/h leak for an AMCO C700 InVision water meter. The range for periodic leaks is 5-200 sec in duration and 10-2000 sec in period. It takes 1-2 hours to reliably detect continuous leaks and 4-6 hours to detect periodic leaks.

## II. DATA

A sampling period of half a second has been selected based on empirical ranges and hardware processing limitations. Data sets for initial exploratory analysis were recorded for a variety of simulated leaks as an overlay to real water consumption patterns. The resulting time series contains about 10 hours of data required for algorithm development. Figure 1 shows representative data sets for continuous and periodic leaks. Distinct low-frequency harmonic signals, and pulses filled with higher frequency harmonic signal, have been observed in both cases.

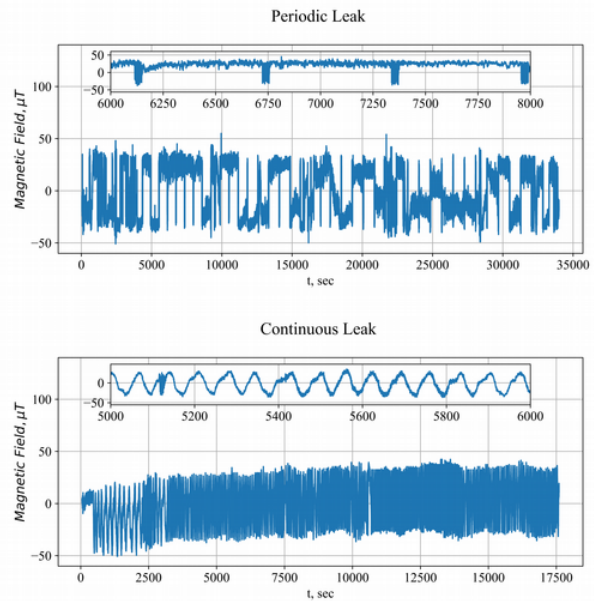

Fig. 1  Examples of continuous and periodic leaks recorded for several hours.

## III. ALGORITHM

Each sample is recorded in a moving window to accumulate statistics for leak detection. Spectra peak frequency and period-duration distributions have been periodically analyzed. The analysis periods that were selected are large enough to average out noise caused by the device and by regular water consumption. The distribution peak significance criteria has been defined after multiple experiments, with special care taken to minimize false positives.

For continuous leaks, spectral peaks of the signal in the entire window are classified and accumulated in the peaks frequency distribution. The spectra is computed every 10 minutes. The distribution analysis is conducted every 2 hours.  A significant

---

The author contact e-mail: agreysukh@gmail.com

peak in the distribution in the expected range indicates a leak. Figure 2 illustrates this case.

For periodic leaks, pulse periods are much lager than a window that would be applicable for using direct spectra analysis. The signal in the beginning of the window gets dereferenced and analyzed to detect the beginning or end of a pulse. A variant of the z-score algorithm, with a 10-sec lag and a threshold of 3, is used to detect pulses. Detected pulses are accumulated in a period-duration distribution. The distribution analysis is conducted every 6 hours. A significant peak in the distribution indicates a leak. Figure 3 illustrates this case.

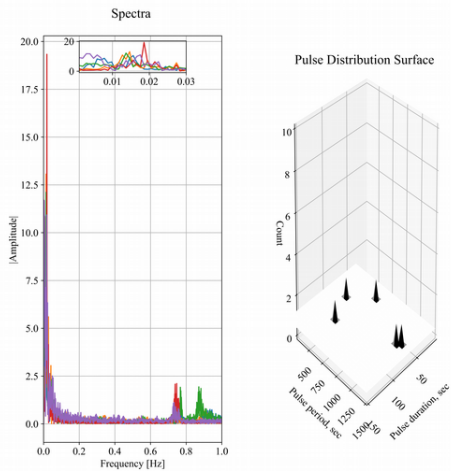

Fig. 2 Continuous leak. Spectra and pulses collected over several hours. Spectra shows distinct peaks in the low-frequency band. The pulse distribution shows no significant peaks.

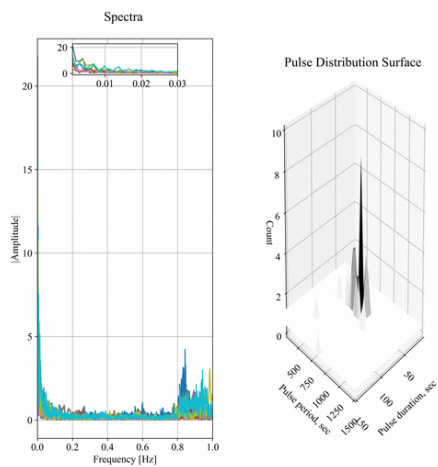

Fig. 3 Periodic leak. Spectra and pulses collected over several hours. Pulse period-duration distribution shows a sharp peak. Spectra has no distinct peaks in the low-frequency band.

## IV. DEVICE

System on chip, ESP32 microcontroller [5], and LSM303 triple-axis magnetometer with an I2C interface [6] is the base of the device. The power source could be any USB module. The controller is powerful enough to conduct all the computations in a fraction of the sample period. Integrated Wi-Fi is used for notifications and system diagnostic.

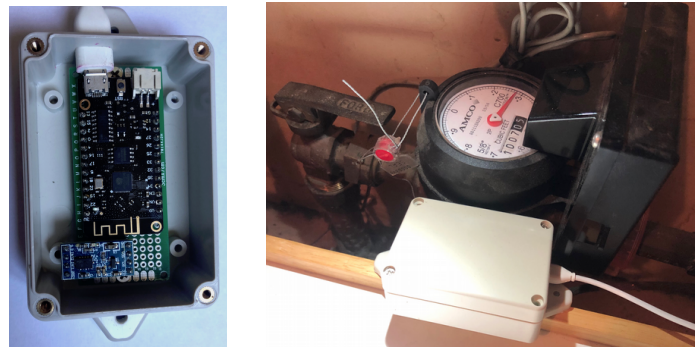

Fig. 4 Micro-controller and magnetometer assembled in a case, placed near the AMCO C700 InVision water meter.

## V. SOFTWARE

Code for the initial exploratory analysis of pre-recorded data is in Python. Preliminary algorithms were also developed with NumPy, pystat, and the Matplotlib Python libraries. After initial development, the software has been optimized and ported to the Arduino C++ environment. A few open-source libraries were utilized, including FFT and Network Time Protocol (NTP) client. Model parameters were tuned on-board during multiple test runs. Special attention has been taken to make the code field-reliable. A simple embedded webserver was used for data monitoring during testing.

## VI. CONCLUSION

The monitor is a functional leak alarm. No false positives were observed during extensive testing. The solution is modular, inexpensive, and focused on a specific problem. Integration with other home automation and resource conservation solutions would be an interesting continuation.

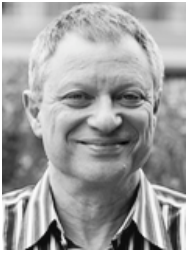
**Alexander Greysukh** received a M.S. degree in engineering physics from St. Petersburg Polytechnic University, Russia.

He has developed scientific and enterprise software for a number of organizations, including NASA JPL, Mitsubishi Electric, and Oracle. Current interests include sensors, automation, signal processing, and data analysis.